\documentclass[sigconf]{acmart}

\usepackage{booktabs} 
\usepackage{fullpage}
\usepackage{amsmath}
\usepackage{graphicx, amssymb}
\usepackage{caption}
\usepackage{subcaption}
\usepackage{amsmath,mathtools}
\usepackage{enumitem}
\usepackage{multirow}
\usepackage{algorithm}
\usepackage{algpseudocode}
\usepackage{makecell}

\setcopyright{rightsretained}

\acmArticle{4}
\acmPrice{15.00}

\begin{document}

\settopmatter{printacmref=false}

\title{An Adjustable Heat Conduction based KNN Approach for Session-based Recommendation}


\author{Huifeng Guo}
\authornote{The work is done when Huifeng Guo works as an intern in Noah's Ark Lab, Huawei.}
\affiliation{%
  \institution{Shenzhen Graduate School, \\Harbin Institute of Technology}
  \city{Shenzhen}
  \country{China}}
\email{huifengguo@yeah.net}

\author{Ruiming Tang}
\affiliation{%
  \institution{Noah's Ark Lab, Huawei}
  \city{Shenzhen}
  \country{China}}
\email{tangruiming@huawei.com}

\author{Yunming Ye}
\authornote{Yunming Ye is the corresponding author.}
\affiliation{%
  \institution{Shenzhen Graduate School, \\Harbin Institute of Technology}
  \city{Shenzhen}
  \country{China}}
\email{yeyunming@hit.edu.cn}

\author{Feng Liu}
\affiliation{%
  \institution{Shenzhen Graduate School, \\Harbin Institute of Technology}
  \city{Shenzhen}
  \country{China}}
\email{liufeng@stmail.hitsz.edu.cn}

\author{Yuzhou Zhang}
\affiliation{%
  \institution{Noah's Ark Lab, Huawei}
  \city{Shenzhen}
  \country{China}}
\email{zhangyuzhou3@huawei.com}

\begin{abstract}
The KNN approach, which is widely used in recommender systems because of its efficiency, robustness and interpretability, is proposed for session-based recommendation recently and outperforms recurrent neural network models. It captures the most recent co-occurrence information of items by considering the interaction time. However, it neglects the co-occurrence information of items in the historical behavior which is interacted earlier and cannot discriminate the impact of items and sessions with different popularity.
Due to these observations, this paper presents a new contextual KNN approach to address these issues for session-based recommendation. Specifically, a diffusion-based similarity method is proposed for considering the popularity of vertices in session-item bipartite network, and a candidate selection method is proposed to capture the items that are co-occurred with different historical clicked items in the same session efficiently. Comprehensive experiments are conducted to demonstrate the effectiveness of our KNN approach over the state-of-the-art KNN approach for session-based recommendation on three benchmark datasets.

\end{abstract}
\keywords{Diffusion model, Session-based Recommendation, Nearest Neighbor}

\maketitle

\section{Introduction}\label{sec:intro}
With the development of Internet and web economic, there have been many web applications, such as online shopping, online news and videos, online social networks, and many more. However, the recommender systems of many applications, particularly those of small retailers, do not track the visit information of all users over a period of time. Moreover, the cookies are also unavailable due to the technology reliability and privacy concerns~\cite{RNNforSession15}.
Even if the visit information of users can be tracked, the number of sessions for a specific user in a small application site is limited and the behavior of users mostly shows session-based traits. Therefore, session-based recommendation, where the task is predicting the next action of a user given the sequence of the actions in the current session, is critical for recommender systems.


Session-based recommendation is a special case of sequential learning, such as basket prediction and music playlist generation. The early approaches are based on the recognition of sequential pattern, which can be used for predicting a user's next action. These approaches are applied in music domain~\cite{ltsp12music} and e-commerce~\cite{SPM12}. While pattern mining approaches are easy to implement and interpretable, its computation complexity is high and it is difficult to find a suitable minimum support threshold. Moreover, in some applications, sequential pattern mining approaches do not lead to better performance compared with KNN approaches~\cite{playlist}. Traditional recommendation methods, such as MF~\cite{mf},
SVD++~\cite{Koren10svd++} for rating, FM~\cite{Rendle12fm}, FFM~\cite{juan16ffm}, DeepFM~\cite{DeepFM} for click prediction, are not well suited for sequential learning. Therefore, several approaches are proposed to capture both sequential information and  personalization recommendation. For instance, FPMC~\cite{RendleFS10FPMC} captures sequential information by order-1 markov chain and personalization by matrix factorization, while FISM~\cite{KabburNK13FISM} captures sequential information (similar to FPMC) and personalization by factorizing the item-item similarity matrix. However, all these methods require \textbf{long-term} user history behavior and usually lead to poor performance when such information is unavailable.

In order to address this limitation, a recurrent neural network, named as GRU4Rec~\cite{RNNforSession15}, is proposed for session-based recommendation. Specifically, GRU4Rec adopts a gated recurrent unit model to fit the sequential information hidden in the session records. Afterwards, two advanced RNN approaches \cite{RNNforSession16} and \cite{ContexNNforSession16} are proposed to incorporate additional item features to achieve higher accuracy. Although RNN is able to learn the transition relationship from the current item to the next item, it still surfers from several limitations: (1) limited ability of capturing co-occurrence information of items; (2) tremendous parameters to trained.

Compared with other methods, the KNN (K Nearest Neighbor) approach is widely used in recommender systems because of its efficiency~\cite{playlist,BellK07SCF,item-cf,unify-cf}, robustness and interpretability. Recently, by taking the contextual information and capturing the co-occurrence information of items into account, a contextual KNN approach (in short, CKNN) is proposed for session-based recommendation. According to the result in~\cite{sessionknn}, CKNN \textbf{outperforms} GRU4Rec~\cite{RNNforSession15}, which models sequential information through the gated recurrent units (GRU), on benchmark datasets. Although CKNN achieves better performance in session-based recommendation, it still has two limitations: (1) not being able to distinguish the influence of items with different popularity; (2) when identifying the relevant sessions of current session, which is the key step of CKNN,  CKNN cannot guarantee the ratio of relevant sessions of different items in the current session, especially the last item, in the relevant session set of the current session.

In order to address these limitations in CKNN approach for session-based recommendation, we propose a new CKNN approach in this paper. The contributions are summarized as follows:
\begin{itemize}
\item To distinguish the importance of different items and sessions when calculating the similarity between sessions, we propose an unified diffusion-based similarity method (DSM).  By introducing two parameters $\lambda$ and $\beta$ to control the impact of session length and item popularity respectively, the similarity between sessions is defined more reasonable. As a result, when adopting DSM with $\lambda=0.5$ and $\beta=0.5$, the performance of CKNN is improved in terms of accuracy and diversity.
\item We propose a candidate selection strategy, namely EPCS, to guarantee the ratio of relevant sessions related to last click of current session in the relevant session set of current session. In addition, we propose a second candidate selection strategy, namely EPCSR, to guarantee the ratio of relevant sessions related different items clicked in current session. Extensive experiments are conducted to evaluate the effectiveness and efficiency of the proposed candidate selection strategies.
\item Incorporating DSM and EPCSR in CKNN approach, we propose our algorithm for session-based recommendation, namely CKNN-DSM-EPCSR. We evaluate our approach on three benchmark datasets, which shows consistent improvement over existing state-of-the-art KNN approach for session-based recommendation.
\end{itemize}

The rest of this paper is organized as follows. Section 2 reviews the related research works in the relevant domains. Section 3 introduces the necessary preliminary of contextual KNN approach for session-based recommendation. Section 4 shows our approach for session-based recommendation, including the main algorithm, diffusion-based similarity method and the candidate selection strategies. We extensively evaluate the effectiveness of our approach in Section 5. Finally, we conclude our work.

\section{Related Work}\label{sec:relatedwork}
Most of the approaches for session-based recommendation are belonging to sequential recommendation.

In this paper, we introduce a new KNN approach for session-based recommendation. The most related domains are session-based recommendation and collaborative filtering for recommendation. In this section, we
discuss related work in these two domains.
\subsection{Session-Based Recommendation}
There are three kinds of approaches in session-based recommendation: pattern-based, sequential-based and neighbor-based technologies.

The pattern-based approaches apply frequent
mining techniques, such as association rules~\cite{AR93} and sequential pattern mining~\cite{SPM12}, to determine the next recommendation. The sequential-based algorithms are divided into Markov model-based approaches and recurrent neural network (RNN)-based approaches. The main assumption of Markov model-based approaches in session-based recommendation is that the selection of next action in a session is dependent only on a limited number of previous actions~\cite{NLPlaylist11,MDP05}. However, the main issue of Markov model-based approaches is the state space becomes unmanageable when all possible sequences of user selection need to be considered~\cite{RNNforSession15}.
Recently, lots of approaches utilize RNN to model session-based recommendation. Hidasi proposes GRU4Rec in~\cite{RNNforSession15} which models the item transactions within sessions by gated recurrent units (GRU). Afterwards, two advanced RNN approaches \cite{RNNforSession16} and \cite{ContexNNforSession16} are proposed which incorporate additional item features to achieve higher accuracy.

Although RNN-based approaches are able to model the sequential transactions within sessions, it is weak in capturing the co-occurrence information of items. Different from finding item-to-item correlations in pattern-based approaches, neighbor-based approaches search for sessions that are similar to the current session and then recommend most similar items in such similar sessions~\cite{sessionknn}. The neighbor approaches are, despite their simplicity, quite effective and efficient.

\subsection{Collaborative Filtering}
The Collaborative Filtering (CF) models are well studied for recommender systems from the last decade \cite{item-cf,unify-cf}. The basic assumption is that users with similar behaviors will like the same kind of items, and the items attracting similar users will share similar ratings from a user. CF-based models have many advantages, such as easy to implement, easy to understand and efficient. Therefore, many methods are studied, including graph-based~\cite{gupta2013wtf, psp2017, push2015}, MF-based~\cite{mf, svd++} and diffusion-based. Among these methods, diffusion-based models are a vital branch based on bipartite or tripartite networks~\cite{NetworkRec15}. Diffusion-based methods simulate a resource-allocation process on bipartite network to make recommendation, which are inspired by the diffusion phenomenon in physical dynamics. The most popular diffusion-based algorithms are mass diffusion (MD)~\cite{Zhou2007Bipartite} and heat conduction (HC)~\cite{Zhang2007Heat}. The former one is equivalent to the basic random walk algorithm where the result is of high accuracy and dominated by the nodes with high degree. In contrast, the latter one usually recommends the result of low accuracy but high diversity. In order to balance the accuracy of MD and the diversity of HC, a hybrid approach is proposed in~\cite{Zhou2010Solving}. In this paper, we investigate a diffusion-based similarity method of KNN approach for session-based recommendation to incorporate the advantages of diffusion-based methods.



\section{Preliminary}\label{sec:algo:pre}

\begin{table}[!t]
\centering
\caption{The example of session-based recommendation.}\label{table:session-item-example}
\begin{tabular}{ccc}
\toprule

 Session id& Item id & time\\
\midrule
i& $\alpha1$ &00 \\
i& $\alpha2$ &01 \\
i& $\alpha3$ &02 \\
i& $\alpha6$ &03 \\
j& $\alpha3$ &04 \\
j& $\alpha4$ &05 \\
k& $\alpha2$ &06 \\
k& $\alpha3$ &07 \\
k& $\alpha4$ &08 \\
l& $\alpha3$ &09 \\
l& $\alpha5$ &10 \\
\bottomrule
\end{tabular}
\end{table}

In order to illustrate the procedure of session-based recommendation, Table~\ref{table:session-item-example}~presents an example data. This example includes 11 session-item interactions, which consists of Session id, Item id and time, respectively. Before introducing the algorithm, we first define several definitions used in this paper.
\begin{definition}
\textbf{Current session} $x$ is the set of clicked items in current session.
\end{definition}
\begin{definition}
The \textbf{Task} of session-based recommendation: based on the given item history of the current session, predict the next action.
\end{definition}
\begin{definition}\label{def:rs}
    \textbf{Relevant session set} $RL(x)$ is a set of sessions, each of which contains at least one item in the current session $x$.
    \begin{displaymath}
    RL(x)=\bigcup_{i \in x}{RL_{i}},
    \end{displaymath}
    where $RL_{i}$ is the set of sessions including item $i$. For example, when current session $x=\{\alpha_4,\alpha_1\}$, the $RL(x)$ is $\{j,k,i\}$,  where $RL_{\alpha_4}=\{j,k\}$ and $RL_{\alpha_1}=\{i\}$ respectively.
\end{definition}
\begin{definition}
    \textbf{Recent relevant session set} $RC(x)$ is the most recent $k_{recent}$ sessions that selected from the relevant session set $RL(x)$, where $\mathbf{k_{recent}}$ is the number of the most recent relevant sessions and specified by users. Following the example in Definition~\ref{def:rs}, the $RC(x)=\{j,k\}$, when we set $k_{recent}=2$.
\end{definition}
\begin{definition}
\textbf{Nearest neighbor session set} $NN(x)$ is the most similar $k_{top}$ sessions that are selected from the recent session set $RC(x)$, where $\mathbf{k_{top}}$ is the number of the most similar recent relevant sessions and specified by users. Note, the similarity between sessions can be calculated by cosine and other similarity methods. Assuming we select cosine as similarity metric and $k_{top}=1$, the $NN(x)=\{j\}$.
\end{definition}

\begin{definition}
The session-item interactions presented in Table~\ref{table:session-item-example} can be represented as a \textbf{session-item bipartite network} $\mathcal{G}=\{\mathcal{S},\mathcal{I},\mathcal{E}\}$, which consists of a session set $\mathcal{S}$, an item
set $\mathcal{I}$ and an interaction set $\mathcal{E}$. The session set is defined as $\mathcal{S}=\{S_1,S_2,\ldots,S_m\}$, the item set is defined as $\mathcal{I}=\{I_1,I_2,\ldots,I_n\}$ and the interaction set is defined as $\mathcal{E}=\{E_1,E_2,\ldots,E_l\}$, where $m$ is the number of sessions, $n$ is the number of items in the session-item interaction set $\mathcal{E}$ and $l$ is the cardinality of $\mathcal{E}$.
\end{definition}

\begin{definition}
 The adjacency matrix of $\mathcal{G}$ can be denoted as $A\in R^{m\times n}$,  where the element of $A$ describes the relationship between the sessions and items:
 \begin{displaymath}
 a_{{x}{i}} =
 \begin{cases}
 1, interaction~(x,i)\in \mathcal{E}, \\
 0, otherwise.
 \end{cases}
 \end{displaymath}
\end{definition}
In complex networks, degree of a vertex is an important concept, which is defined as the number of edges linked to this vertex. Accordingly, we define the degree of a session $x$ and the degree of item $i$ as $\mathbf d_x=\sum_{i=1}^{n}a_{{x}{i}}$ and $\mathbf  d_i=\sum_{j=1}^{m}a_{{j}{i}}$, which represent the number of items in session $x$ and the times that the item $i$ appearing in a session, respectively.

\section{Our Approach}\label{sec:algo}

In this section, the procedure of the contextual KNN (CKNN for short) approach for session-based recommendation is presented at first. Then, to capture more co-occurrence information of items and incorporate the graph structure information into the KNN approach,
the candidate selection and the diffusion-based similarity methods are introduced in Section~\ref{sec:algo:cs} and Section~\ref{sec:algo:sim}, respectively.

\subsection{Contextual KNN Approach for Session-based Recommendation}\label{sec:algo:cknn-overview}

\begin{figure}
\centering
\includegraphics[width=0.5\textwidth]{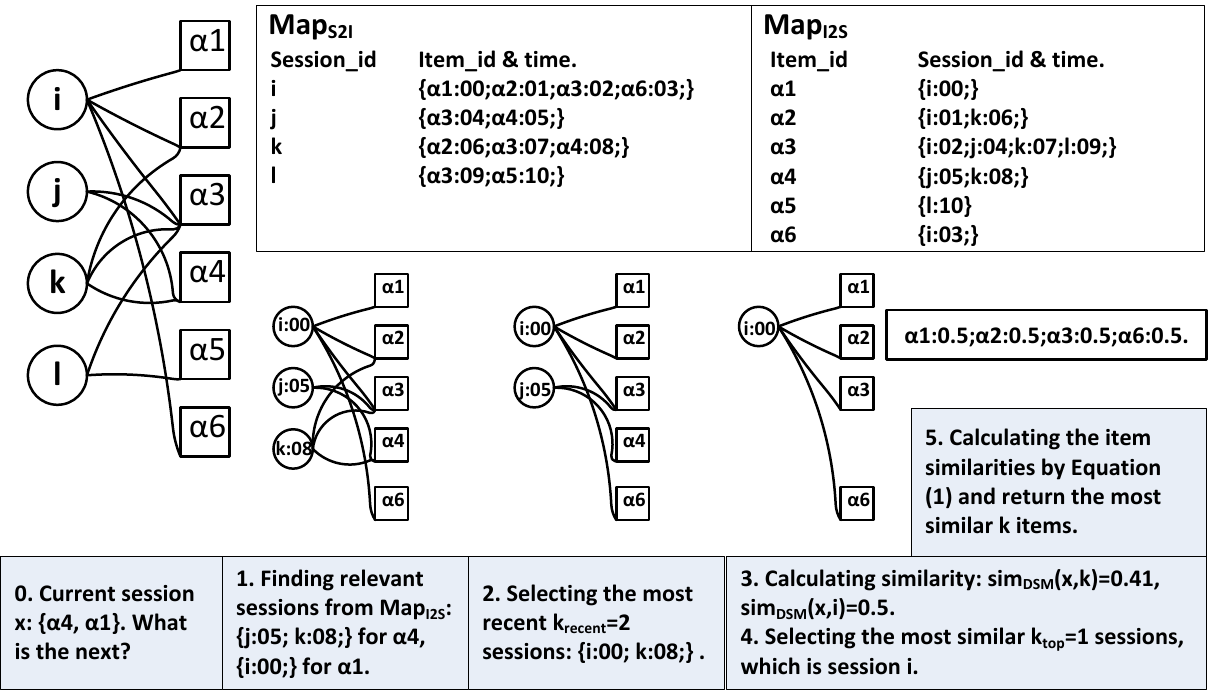}
\caption{\small{Examples for bipartite network construction and our session-based KNN algorithm. Note that, $Map_{S2I}$ (or $Map_{I2S}$) is a map, where the key is session id (or item id) and the value is the set of item id (or session id)-timestamp pairs. Specifically, the smaller the timestamp, the earlier the interaction time is (03 is more recent than 00).}}\label{fig:example}
\end{figure}

 Before staring, we need construct two dictionaries, namely $Map_{S2I}$ and $Map_{I2S}$. Specifically, $Map_{S2I}$ is a map from session to the pairs of item and interaction time, $Map_{I2S}$ is another map from item to the pairs of session and interaction time. The $Map_{S2I}$ and $Map_{I2S}$ of the example data that introduced in Table~\ref{table:session-item-example} are presented in Table~\ref{table:session:example-s2i} and Table~\ref{table:session:example-i2s}, respectively. The pipeline of CKNN approach is described as follows (which is also presented in the bottom of Figure~\ref{fig:example}):

\begin{table}[!t]
\centering
\caption{The map from session id to the pair of item id and interaction time.}\label{table:session:example-s2i}
\begin{tabular}{cc}
\toprule
 Session id& (Item id, interaction time)\\
\midrule
i& ($\alpha1$, 00), ($\alpha2$, 01), ($\alpha3$, 02), ($\alpha6$, 03) \\
j& ($\alpha3$, 04), ($\alpha4$, 05)\\
k& ($\alpha2$, 06), ($\alpha3$, 07), ($\alpha4$, 08)\\
l& ($\alpha3$, 09), ($\alpha5$, 10)\\
\bottomrule
\end{tabular}
\end{table}

\begin{table}[!t]
\centering
\caption{The map from item id to the pair of session id and interaction time.}\label{table:session:example-i2s}
\begin{tabular}{ccc}
\toprule

 Item id& (Session id, interaction time)\\
\midrule
$\alpha1$ &(i, 00) \\
$\alpha2$ &(i, 01), (k, 06)\\
$\alpha3$ &(i, 02), (j, 04), (k, 07), (l, 09) \\
$\alpha4$ &(j, 05), (k, 08) \\
$\alpha5$ &(l, 10) \\
$\alpha6$ &(i, 03) \\
\bottomrule
\end{tabular}
\end{table}

\begin{itemize}
\item \textbf{At step 0}, a session-based recommendation is triggered when a user clicks some item. \textbf{At step 1}, we need to find \textbf{RL(x)}, which is the relevant session set related to items in the current session $x$. For example in Figure~\ref{fig:example}, the relevant session set related to the current session $x=\{\alpha_4,\alpha_1\}$ is $\{j,k,i\}$ (session $j,k$ are related to item $\alpha_4$ and session $i$ is related to item $\alpha_1$).

\item \textbf{At step 2}, we select the most recent $k_{recent}$ sessions from the relevant session set as the recent session set \textbf{RC(x)}, because focusing on the most recent events has shown to be effective in the domains of e-commerce and news recommendations \cite{DeepYoutube16}. For example in Figure~\ref{fig:example}, $k_{recent}=2$ recent sessions are chosen and the recent session set of current session $x$ is $\{j,k\}$.


\item \textbf{At step 3}, the similarity score $sim(x,j)$  between the current session $x$ and each session $j$ in RC(x) (selected by step 2) is calculated. After that, the most similar $k_{top}$ sessions are selected as the nearest neighbor session set, denoted as $NN(x)$, for current session $x$.
\item \textbf{Finally at step 4}, based on the nearest neighbor session set $NN(x)$, and the similarities with the current session $x$, the score of a recommendable item $\alpha$ for the current session $x$ is
    \begin{equation}\label{eq:sim}
    score_{KNN}(\alpha,x)=\sum_{j\in NN(x)} sim(x,j)\times 1_j(\alpha),
    \end{equation}
    where $1_j(\alpha)=1$ represents session $j$ containing item $\alpha$ and $0$ otherwise. Then the most similar $k$ items $result_x$ are returned.
\end{itemize}


%

\subsection{Candidate Selection}\label{sec:algo:cs}
Focusing on the most recent events has shown to be effective in the domains of e-commerce and news recommendation~\cite{DeepYoutube16}. Therefore, \cite{sessionknn} proposes an algorithm, denoted as \textbf{Original} for distinction, to select the most $k_{recent}$ recent relevant sessions from the related session set $RL(x)$ of current session $x$. However, this algorithm still surfers following limitations:

\begin{itemize}
\item \textbf{Last click}: The items are interacted by users sequentially in the scenario of session-based recommendation, therefore, the most recently interacted item is critical for the next item recommendation. the \textbf{Original} algorithm proposed by \cite{sessionknn} ignores such information. Specifically, the sessions containing the last item may not be included because the interaction time of such sessions may be earlier than the other selected sessions. As a result, it is not guaranteed to include the relevant sessions of last click.
\item \textbf{Other clicks}: Same as \textbf{Last click}, the relevant sessions for other clicks will be excluded when the interaction time are earlier.

\end{itemize}
In the remainder of this section, we introduce different strategies to handle these limitations.
\subsubsection{Focusing on the Last Click}\label{sec:algo:cs:flc}
In order to focus on the last click item, we propose a strategy to guarantee the ratio of last item's recent relevant sessions in the recent session set $RC$. The basic idea of our strategy is to find the recent session set from two sources: (1) the recent relevant session set of the last item; (2) the recent relevant session set of the other items.
    \begin{itemize}
    \item \textbf{Recent relevant sessions for last click}: Assuming item $i$ is the last item, we select the most recent
    ${ [{k_{recent}}\times p_{i}]}$\footnote{[~] indicates rounding.} sessions from $RL_{i}$ as $RC_{i}$, where $\mathbf p_{i}$ is the ratio of $RC_{i}$ in $RC(x)$ according to different strategies.
    \item \textbf{Recent relevant sessions for other clicks}: Selecting the most recent $[k_{recent}\times (1-p_{i})]$ sessions of  $x\setminus i$ (other items in current session) as $RC(x\setminus i)$ from $RL(x\setminus i)$, which has already been found previously.
    \item \textbf{Recent relevant sessions for current session}: Merging the recent session set of last item and of the other items in current session $x$ as the recent session set of $x$:
        \begin{displaymath}
         RC(x)=RC_{i}\cup RC(x\setminus i).
         \end{displaymath}
    \end{itemize}

Assuming the influence of items with different popularity is same, we propose \textbf{E}qual \textbf{P}robability \textbf{C}andidate \textbf{S}election (EPCS for short). So $\mathbf p_{i}=[\frac{k_{recent}}{|x|}]$, where $|x|$ is the number of elements in current session $x$.

To make it easier to understand, we take an example in Table~\ref{table:candidate} to illustrate the difference between the \textbf{Original} algorithm proposed by \cite{sessionknn} and the \textbf{EPCS} algorithm in this paper. Recall the items of current session $x$ are $\{\alpha_4,\alpha_1\}$ (the same as Figure~\ref{fig:example}), where $\alpha_1$ is the last interacted item. Therefore, the $RL(x)$ is $\{j:05, k:08, i:00\}$ (where $RL_{\alpha_4}=\{j,k\}$ and $RL_{\alpha_1}=\{i\}$). If we set $k_{recent}=2$, session $i$ will be excluded because the interaction time of $\{i:00\}$ is in the third place. Therefore, there is no session relevant to last item $\alpha_1$ in the recent session set. On the contrary, EPCS \textbf{guarantees the ratio} of $RC_{i}$ in $RC(x)$.

\begin{table}[ht]
\centering
\caption{Example for candidate selection.}\label{table:candidate}
\begin{tabular}{cccccc}
\toprule
   &  candidate sessions & recent sessions \\
  \midrule
 \cite{sessionknn} & $\{j:05, k:08, i:00\}$ & $\{k:08, j:05\}$ \\
Focussing on Last Click& $\{j:05, k:08, i:00\}$ & $\{k:08, i:00\}$  \\
  \bottomrule
\end{tabular}
\end{table}

\subsubsection{Focusing on All Clicks}\label{sec:algo:cs:fac}
In fact, not only the last click is important, but also the co-occurrence of other clicks is helpful for recommendation. To guarantee the ratio of every item's recent relevant sessions in $RC(x)$, we can select the recent session set for each item that included in current session, then aggregate these sets together as the recent session set of current session $x$. However, selecting the recent session set for each item is very expensive in terms of time cost, because querying and sorting are needed every time.

Recall the candidate selection algorithm introduced in Section~\ref{sec:algo:cs:flc}, the recent session set of other clicks ($x\setminus i$) is selected at previous step. So we can select $[k_{recent}\times (1-p_{i})]$ elements from ($x\setminus i$) randomly. This approach has \textbf{two advantages}: (1) Random selection is efficient; (2) Random selection guarantees the ratio of every item's relevant sessions in $RC(x)$. Based on the \textbf{random} selection strategy and \textbf{EPCS}, we propose another strategy and name it as \textbf{EPCSR}.

\subsection{Diffusion-based Similarity}\label{sec:algo:sim}

In addition to candidate selection, similarity calculation is also important for the performance of CKNN approach. For the purpose of incorporating with the graph structure in the procedure of similarity calculation (step 3 as shown in Figure~\ref{fig:example}), we propose the diffusion-based similarity method for session-based recommendation. In this section, we first review several popular diffusion models in recommender systems in Section~\ref{sec:algo:sim:dm}, then, on the basis of the existing diffusion models, we propose our diffusion-based similarity method in Section~\ref{sec:algo:sim:adhc}.

%
%
%

\subsubsection{Diffusion Model}\label{sec:algo:sim:dm}
The diffusion-based process can be described as a two-step resource-allocation process on a bipartite network. Specifically, the mass diffusion (MD~\cite{Zhou2007Bipartite}) and heat conduction (HC~\cite{Zhang2007Heat}) are widely used in recommender system~\cite{Zhou2010Solving} because of their simplicity and efficiency. The MD algorithm is equivalent to the random walk where the result is of high accuracy, while HC is proposed to address the challenge of diversity~\cite{Zhou2010Solving}.

Different from traditional collaborative filtering, we only need the first step to compute the similarity between sessions in the scenario of session-based recommendation. In the following, we describe the process of calculating the MD-based and HC-based similarity on the session-item bipartite network.
\begin{itemize}
\item \textbf{MD-based similarity: }At the beginning, items associated with the current session $x$ on the bipartite network obtain resource from $x$ (whose initial resource $c_x$ is 1):
    \begin{displaymath}
    c_{i}=a_{{x}{i}}\frac{c_x}{d_{x}}.
    \end{displaymath}
     Then, the resource of item flows to neighboring sessions based on each item's degree:
\begin{displaymath}
\hat{c}_j=\sum_{i=0}^{n}a_{{j}{i}}\frac{c_{i}}{d_i}=
\sum_{i=0}^{n}a_{{j}{i}}\frac{a_{{x}{i}}\frac{c_{x}}{d_x}}{d_i},
\end{displaymath}
where $j$ represents one of the target sessions which will get resource from the current session $x$; $c_{i}$ and $\hat{c}_j$ are the resource on item $i$ and session $j$, respectively. So the MD-based similarity between current session $x$ and the target session $j$ is:
\begin{displaymath}
Sim_{MD}(x,j)=\frac{\hat{c}_j}{c_x}=\sum_{i=0}^{n}\frac{a_{{x}{i}}\cdot a_{{j}{i}}}{{d_x}\cdot {d_i}}.
\end{displaymath}


\item \textbf{HC-based similarity:} Compared to MD, HC is a bit different. The basic assumption of HC is that the temperature of each item is equivalent to the average temperature of the related sessions~\cite{Zhang2007Heat}. For example, suppose the initial temperature of the current session $x$ is 1 (while the initial temperature of other sessions and items are all zero), then the temperature of item $i$ is the average temperature of the sessions which contain item $i$:
    \begin{displaymath}
     c_{i}=\frac{a_{{x}{i}}\cdot {c_{x}}}{d_i}.
    \end{displaymath}
     Then the temperature is conducted to neighboring sessions based on each session's degree:
\begin{displaymath}
\hat{c}_j=\sum_{i=0}^{n}\frac{a_{{j}{i}}\cdot{c_{i}}}{K(s_{j})}=
\sum_{i=0}^{n}\frac{a_{{j}{i}}\cdot{\frac{a_{{x}{i}}\cdot {c_{x}}}{d_i}}}{d_j}.
\end{displaymath}
So the HC-based similarity between the current session $x$ and target session $j$ is:
\begin{equation}
Sim_{HC}(x,j)=\frac{\hat{c}_j}{c_x}=\sum_{i=0}^{n}\frac{a_{{j}{i}}\cdot{a_{{x}{i}}}}{d_i\cdot d_j}.
\end{equation}

\item \textbf{MDHC-based similarity:} To take the advantages of high accuracy of MD and good diversity of HC, \cite{Zhou2010Solving} proposes a hybrid method:
    \begin{equation}
Sim_{MDHC}(s,j)= \frac{1}{{{d_x}^{\lambda}}\cdot{{d_j}^{1-\lambda}}}\sum_{i=0}^{n}\frac{a_{{j}{i}}\cdot{a_{{x}{i}}}}{d_i},
\end{equation}
where $\lambda$ is a hyper-parameter to balance the MD and HC methods. The hybrid method is exactly MD when $\lambda=1$ and is equivalent to HC when $\lambda=0$.
\end{itemize}

\subsubsection{Diffusion-based Similarity Method}\label{sec:algo:sim:adhc}
Though the HC, MD and MDHC similarity consider the graph structure, there is still a limitation. In CKNN algorithm, some interactions are discarded when constructing the recent session set of current session, which leads to the fact that the degree of items in recent session set is smaller than their real degree in the original network $\mathcal{G}$.
As a result, using items' original degree will over-affect the similarity metric. To address this limitation, we propose a \textbf{D}iffusion-based \textbf{S}imilarity \textbf{M}ethod (noted as DSM for short). In DSM, we adopt an exponential function of items' degree to control the impact of items' popularity for similarity calculation. Specifically, the importance of item $i$ is denoted as ${d_i}^{\beta}$, where $\beta$ is a hyper-parameter and the importance of item degree is increasing when $\beta$ becomes larger.
%
The DSM similarity between current session $x$ and target session $j$ is denoted as:
\begin{equation}
\label{eq:AHCM}
Sim_{DSM}(x,j,\lambda, \beta)=\frac{1}{{d_x}^{\lambda}\times {d_j}^{1-\lambda}}\sum_{i=0}^{n}\frac{a_{{x}{i}}\cdot a_{{j}{i}}}{{d_i}^{\beta}}.
\end{equation}

\begin{table}[ht]
\centering
\caption{Relationship between DSM and other different similarity methods.}\label{table:relationship_sim}
\begin{tabular}{cccccc}
\toprule
  Parameter &cosine&MD&HC&MDHC\\
  \midrule
$\beta$ &0      &1  &1  &1\\
$\lambda$ &0.5  &1  &0  &[0, 1]\\
  \bottomrule
\end{tabular}
\end{table}
It's easy to find that DSM is a general framework over several existing similarity methods. As shown in Table~\ref{table:relationship_sim}, we can obtain different similarity methods by adjusting $\lambda$ and $\beta$. In addition to MD, HC and MDHC, DSM is equivalent to cosine similarity when $\lambda=0.5$ and $\beta=0$:
\begin{displaymath}
Sim_{cosine}(x,j)=\frac{\sum_{\alpha=0}^{n}{a_{x\alpha}\cdot a_{j\alpha}}}{\sqrt{K(s_{x})}\cdot\sqrt{K(s_{j})}}
=Sim_{DSM}(x,j,0.5, 0).
\end{displaymath}

Furthermore, the influence of the current session's length is same when comparing the similarity values between current session $x$ and the recent sessions:
\begin{displaymath}
\frac{Sim_{DSM}(x,j)}{Sim_{DSM}(x,k)}=\frac{\frac{1}{{{d_x}^{\lambda}}\cdot{{d_j}^{1-\lambda}}}\sum_{i=0}^{n}\frac{a_{{j}{i}}\cdot{a_{{x}{i}}}}{{d_i}^{\beta}}}
{\frac{1}{{{d_x}^{\lambda}}\cdot{{d_k}^{1-\lambda}}}\sum_{i=0}^{n}\frac{a_{{k}{i}}\cdot{a_{{x}{i}}}}{{d_i}^{\beta}}}=\frac{\sum_{i=0}^{n}\frac{a_{{x}{i}}\cdot a_{{j}{i}}}{{d_i}^{\beta}\cdot {d_j}^{1-\lambda}}}{\sum_{i=0}^{n}\frac{a_{{x}{i}}\cdot a_{{k}{i}}}{{d_i}^{\beta}\cdot {{d_k}^{1-\lambda}}}}.
\end{displaymath}
Therefore, the DSM  for session-based recommendation is simplified to:
\begin{equation}
\label{eq:AHCM}
Sim_{DSM}(x,j,\lambda, \beta)=\frac{1}{{d_j}^{1-\lambda}}\sum_{i=0}^{n}\frac{a_{{x}{i}}\cdot a_{{j}{i}}}{{d_i}^{\beta}}.
\end{equation}


%


\section{Experiment}\label{sec:expe}
There are two key contributions of this work for session-based recommendation: (1) candidate selection strategies are designed for focusing last click and all clicks; (2) we propose a diffusion-based similarity method for session-based recommendation. In this section, we conduct experiments to answer the following research questions:
\begin{itemize}[leftmargin=1em]
\item \textbf{RQ1: } Does the candidate selection strategies (i.e., EPCS and EPCSR) improve the performance of session-based recommendation?
\item \textbf{RQ2: } How do the hyper-parameters $\lambda$ and $\beta$ in DSM influence the session-based recommendation?
\item \textbf{RQ3: } How does our overall approach perform, compared to the state-of-the-art KNN approaches for session-based recommendation?
\end{itemize}

In what follows, we first present the experimental settings, followed by answering the above research questions one by one.
\subsection{Experiment Setting}

\subsubsection{Datasets}
The effectiveness of our proposed approach is evaluated on three datasets, including RSC, RSCW and Atom. The recommendation task is to predict the $k^{th}$ item knowing the previous $k-1$ items in a session with length $n$, where $k\in [2,n]$.

Particularly, Atom is a dataset containing music playlists from artofthemix.org. The playlists in Atom dataset have no timestamp information therefore the authors of \cite{sessionknn} assigns each playlist with a timestamp uniformly at random, under the assumption that the whole dataset is of 31 consecutive days. The training set of Atom is the first 30 days' data while the last day's data is the test set.

RSC and RSCW are 2 variants of the ACM RecSys 2015 Challenge dataset (RSC15) as used in \cite{sessionknn}. RSC15 contains the sessions of items in 182 consecutive days. Note that, all of the session-item interactions in RSC and RSCW are associated with timestamps.
In RSC, the first 181 days' data is identified as training set, while the last day's data is left as test set. And the RSCW dataset is constructed by selecting five subsets of 91 consecutive days' data from the original RSC15. In each of such subsets, the first 90 days' data is the training set, while the last day's data is the test set. Hence, RSCW dataset contains five sub-datasets, in each of which the recommendation problem is set as in RSC dataset. The performance on RSCW is averaged on the five sub-datasets, to minimize the risk that the obtained results are sensitive to the train-test splitting strategy.

The statistic information of the three datasets are shown in Table~\ref{table:dataset}. The statistics of RSCW is averaged on its five sub-datasets.

\begin{table}[ht]
\centering
\caption{\small{Dataset characteristics.}}\label{table:dataset}
\begin{tabular}{cccccc}
\toprule
   & RSC & RSCW & Atom \\
  \midrule
 Sessions & 8M & 4M & 82K \\
 Avg. Length&3.97 & 3.92 &11.48  \\
 Items & 37K &  34K&  54K\\
 With Timestamp &Yes&Yes& No\\
  \bottomrule
\end{tabular}
\end{table}
\subsubsection{Metrics}
In our experiments, we adopt three evaluation metrics: \textbf{Hit Rate (HR)}, \textbf{Mean Reciprocal Rank (MRR)} and \textbf{Coverage}. In the following, the test dataset is denoted as $Test$, the  cardinality of the test dataset is denoted as $|Test|$, $R_i@L$ is the recommendation list at length $L=20$ for current session $i$, $I$ is the item set and its cardinality is denoted as $|I|$.

\begin{equation}
 HR@L=\frac{1}{|Test|}\sum_{i \in Test} hit_i,
 \end{equation}
where $hit_i=1$ when the ground truth item of current session $i$ is recommended in $R_i@L$.
 \begin{equation}
MRR@L=\frac{1}{|Test|}(\sum_{j\in Test}\frac{1}{rank_{j}}),
 \end{equation}
where $rank_j$ indicates the rank of item which is interacted in sample $j$.
MRR is a widely used metric in information retrieval, where the rank of the ground truth item in $R_i@L$ is valued.
 \begin{equation}
Coverage@L=\frac{1}{|I|}|{\bigcup_{j\in Test}{R_j@L}}|.\\
\end{equation}
The coverage describes the  percentage of recommended items in Top-${L}$ places of all the samples' recommendation lists over all the candidate items.

\subsubsection{Compared Approaches}
We conduct experiments to compare the following approaches:
\begin{itemize}
\item \textbf{IKNN}~\cite{RNNforSession15} proceeds in an item-centric manner. In IKNN, the most similar items of current item are selected through an item-item similarity matrix, which has been established based on session-item records.
\item \textbf{CKNN-cosine-Original}~\cite{sessionknn} proceeds in an session-centric manner. Specifically, it adopts cosine as the similarity metric and \textbf{Original} as the candidate selection method. It is the state-of-the-art KNN approach for session-based recommendation.
\item \textbf{CKNN-\{MD, HC, MDHC, DSM\}-Original}: To compare the performance of different diffusion-based similarity methods, we conduct the experiments of CKNN approaches equipped with MD, HC, MDHC and DSM similarity metrics, and the \textbf{Original} candidate selection method.

\item \textbf{CKNN-\{cosine, DSM\}-\{EPCS, EPCSR\}}: To compare different candidate selection strategies, we conduct the experiments of  CKNN approaches equipped with EPCS and EPCSR under two similarity metrics, namely cosine and DSM.
\end{itemize}

For the approaches of CKNN, we set the number of the most recent sessions $k_{recent}=1000$ and the number of the most similar sessions $k_{top}=500$ according to the parameters which achieve the best performance in~\cite{sessionknn}.

\subsection{The Performance of Candidate Selection Strategy (RQ1)}\label{sec:expe:TBCS}
In this section, we present two experiments to evaluate the effectiveness and efficiency of the different candidate selection strategies in Section~\ref{sec:expe:TBCS:effectiveness} and Section~\ref{sec:expe:TBCS:efficiency}, respectively. Noted that, both $\lambda$ and $\beta$ in DSM are set as 0.5.
\subsubsection{The effectiveness of Candidate Selection Strategy}\label{sec:expe:TBCS:effectiveness}
\begin{figure*}[!t]
\centering
\begin{minipage}{0.3\textwidth}
\includegraphics[width=1\textwidth]{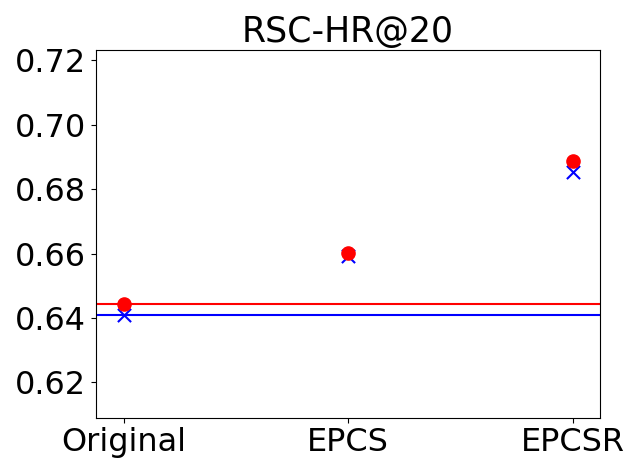}
\subcaption{}
\label{fig:RSC-HR20}
\end{minipage}
\begin{minipage}{0.3\textwidth}
\label{fig:RSC-MRR20}
\includegraphics[width=1\textwidth]{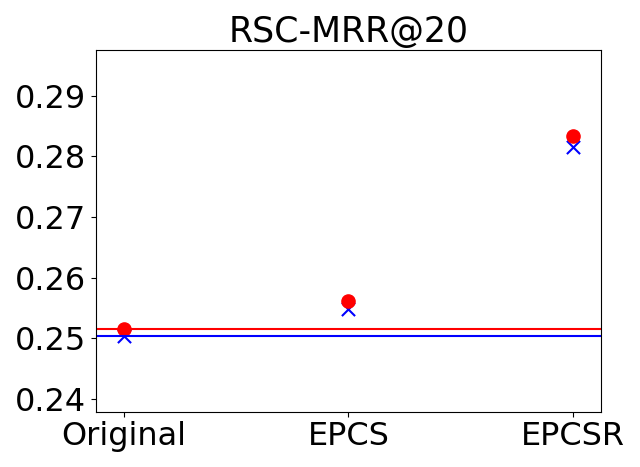}
\subcaption{}
\end{minipage}
\begin{minipage}{0.3\textwidth}
\label{fig:RSC-Coverage20}
\includegraphics[width=1\textwidth]{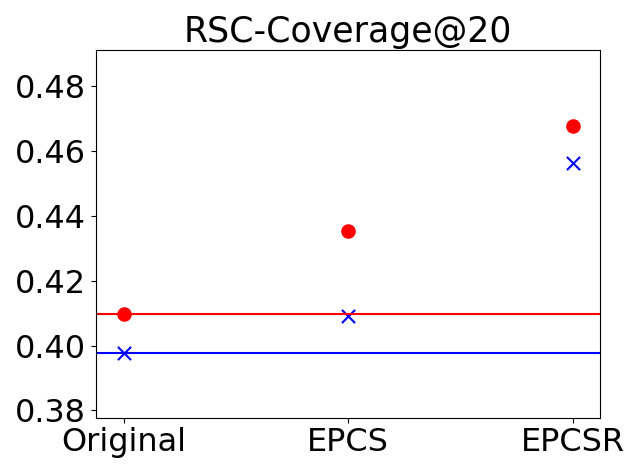}
\subcaption{}
\end{minipage}
\\
\begin{minipage}{0.3\textwidth}
\label{fig:RSCW-HR20}
\includegraphics[width=1\textwidth]{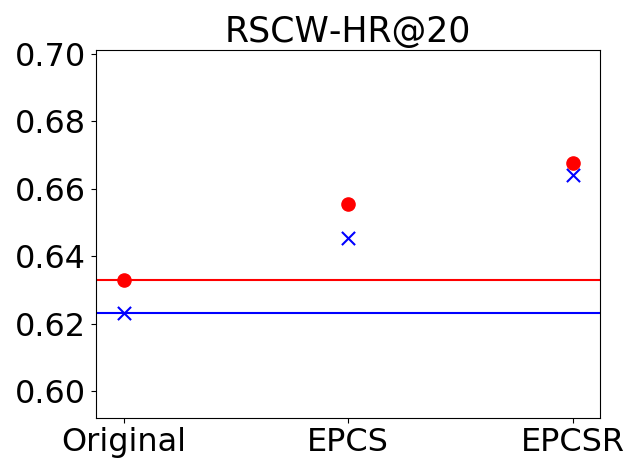}
\subcaption{}
\end{minipage}
\begin{minipage}{0.3\textwidth}
\label{fig:RSCW-MRR20}
\includegraphics[width=1\textwidth]{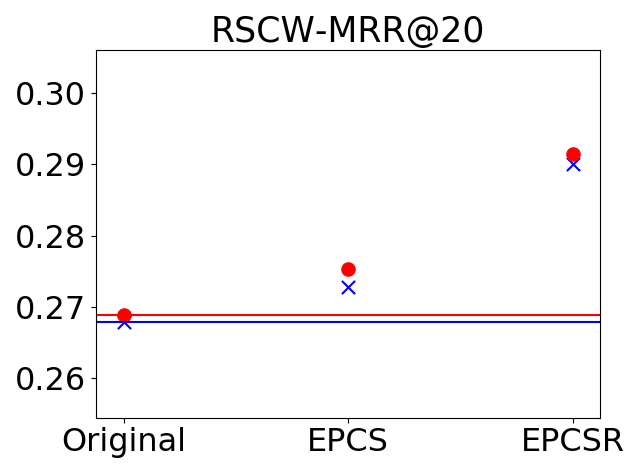}
\subcaption{}
\end{minipage}
\begin{minipage}{0.3\textwidth}
\label{fig:RSCW-Coverage20}
\includegraphics[width=1\textwidth]{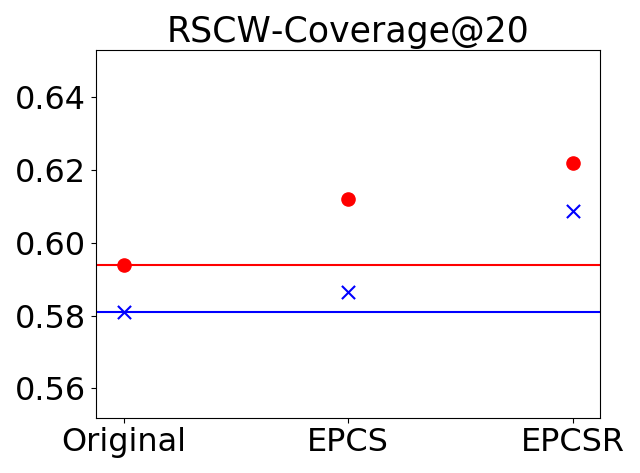}
\subcaption{}
\end{minipage}
\\
\begin{minipage}{0.3\textwidth}
\label{fig:Atom-HR20}
\includegraphics[width=1\textwidth]{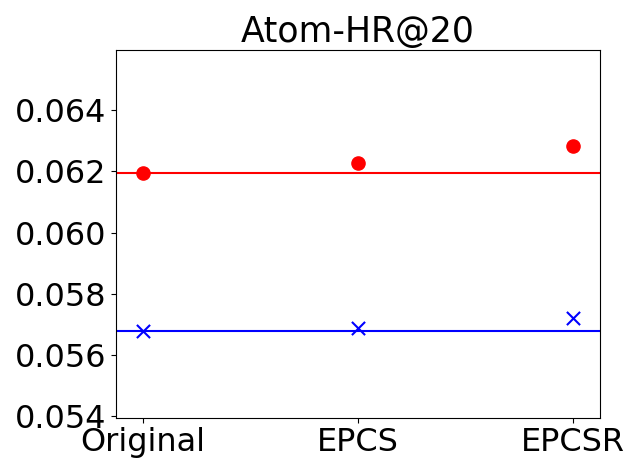}
\subcaption{}
\end{minipage}
\begin{minipage}{0.3\textwidth}
\label{fig:Atom-MRR20}
\includegraphics[width=1\textwidth]{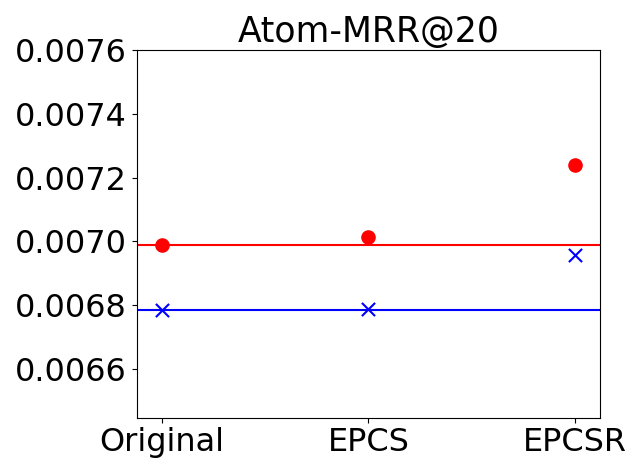}
\subcaption{}
\end{minipage}
\begin{minipage}{0.3\textwidth}
\label{fig:Atom-Coverage20}
\includegraphics[width=1\textwidth]{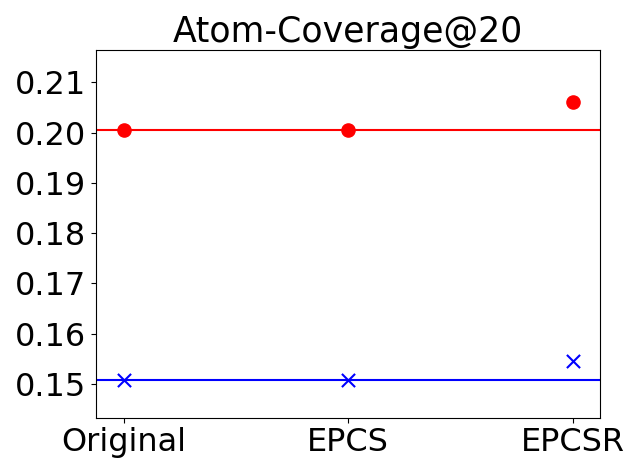}
\subcaption{}
\end{minipage}
\caption{\small{The performance of different candidate selection strategies. Note: the red solid circle and blue $\times$ represent the result of CKNN algorithm when using DSM(0.5, 0.5) and cosine respectively. }}\label{fig:candidate_improve}
\end{figure*}

Figure~\ref{fig:candidate_improve} show the performance of CKNN-DSM(0.5, 0.5) and CKNN-cosine under different candidate selection strategies in terms of HR@20, MRR@20 and Coverage@20 on RSC, RSCW and Atom datasets, respectively. Specifically, the candidate selection strategies include Original, EPCS and EPCSR. The observations can be summarized as follows:
\begin{itemize}
\item Focusing on last click is able to enhance the performance of session-based recommendation with timestamp. It can be observed that the accuracy and diversity of both DSM(0.5, 0.5) and cosine are improved on RSC and RSCW datasets when we adopt EPCS. The reason for this improvement is that the EPCS algorithm guarantees the ratio of relevant sessions containing the last item in the relevant session set, so that the items co-occurred with the last item in the current session can be captured.

\item EPCSR is able to focus on all clicks in current session $x$ and guarantee the ratio of every click's recent relevant sessions in $RC(x)$. Therefore, on the basis of EPCS, EPCSR further enhances the performance and achieves the best results on three datasets.
\end{itemize}

\subsubsection{The efficiency of Candidate Selection Strategy}\label{sec:expe:TBCS:efficiency}
The EPCS and EPCSR is more efficient than the Original approach \cite{sessionknn}. The reasons are as follows: (1) the recent relevant session set of the other items in current session has been found previously, only slightly extra calculation is needed; (2) selecting the recent session set of a single item (i.e., the last item in the session) is much easier and requires less computation than finding that of a set of items.

\begin{figure}[!t]
\centering
\includegraphics[width=0.4\textwidth]{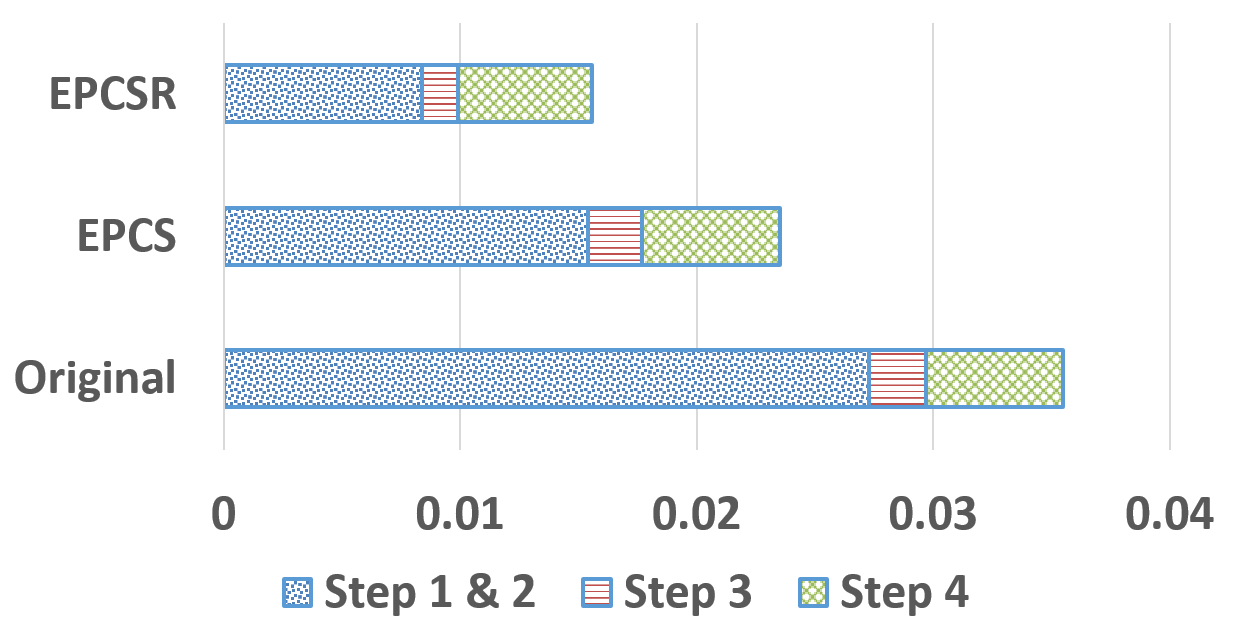}
\caption{\small{The running time comparison of different candidate selection strategies.}}\label{fig:cs-time}
\end{figure}

To summarize, the recent session set of the current session is constructed incrementally by our strategy. Figure~\ref{fig:cs-time}~presents the running time comparison between different candidate selection strategies on RSC dataset, where the x axis represents the running time in seconds. The efficiency of these three strategies is ranked as follows: EPCSR$>$EPCS$>$Original.
\subsection{The Influence of Hyper-Parameters in DSM (RQ2)}\label{sec:expe:HyperPara}
\begin{figure*}[!t]
\centering
\includegraphics[width=1\textwidth]{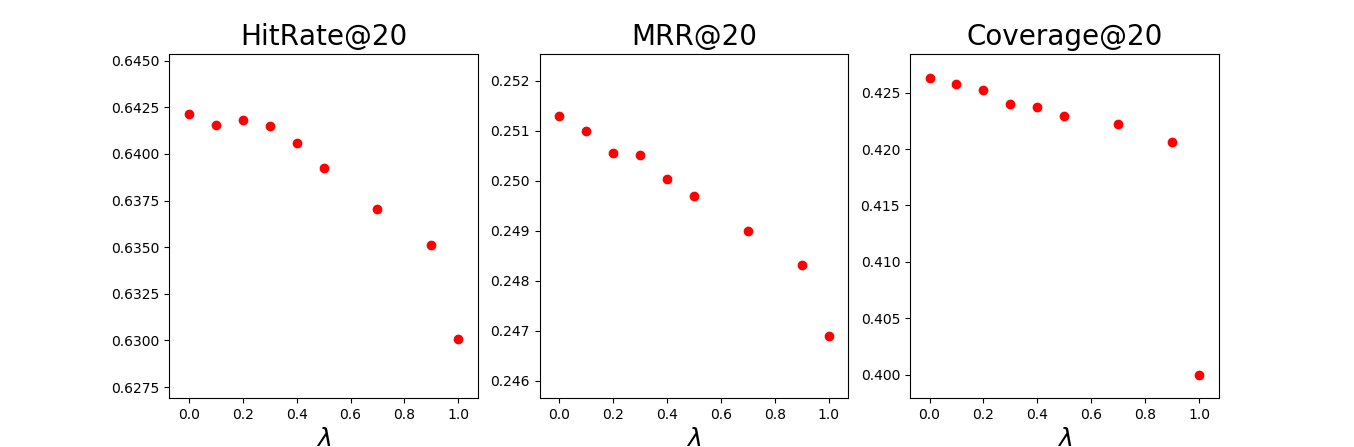}
\caption{\small{The impact of Hyper-Parameter $\lambda$ when $\beta=1$.}}
\label{fig:hyperparameter-lambda}
\end{figure*}

\begin{figure*}[!t]
\centering
\includegraphics[width=1\textwidth]{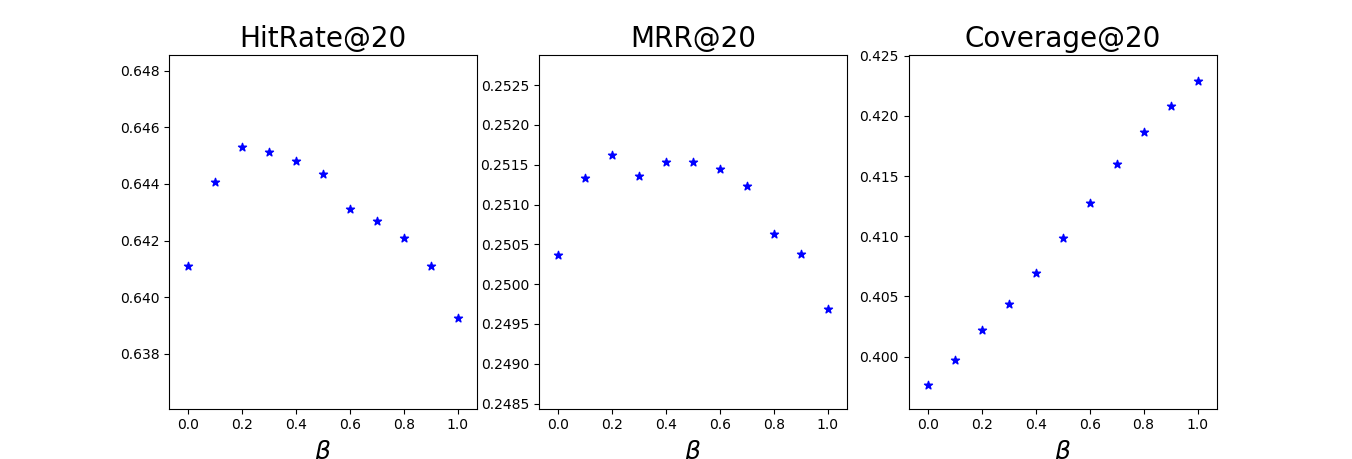}
\caption{\small{The impact of Hyper-Parameter $\beta$ when $\lambda=0.5$.}}
\label{fig:hyperparameter-beta}
\end{figure*}

\begin{figure*}[!t]
\centering
\begin{minipage}{0.3\textwidth}
\label{fig:hr20}
\includegraphics[width=1\textwidth]{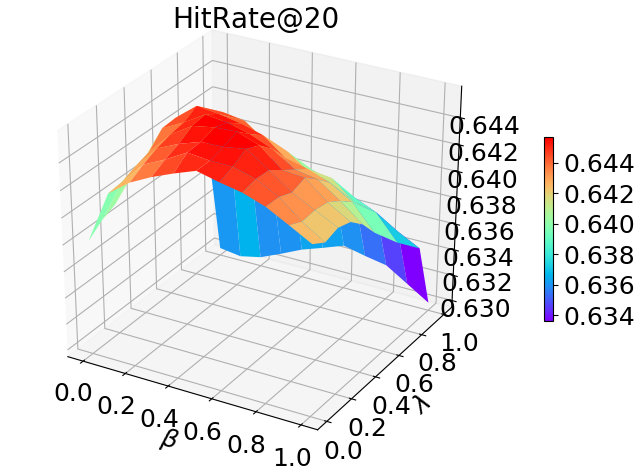}
\end{minipage}
\begin{minipage}{0.3\textwidth}
\label{fig:mrr20}
\includegraphics[width=1\textwidth]{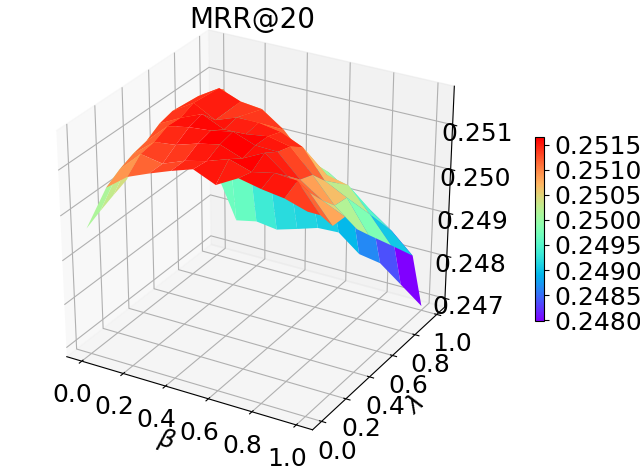}
\end{minipage}
\begin{minipage}{0.3\textwidth}
\label{fig:coverage20}
\includegraphics[width=1\textwidth]{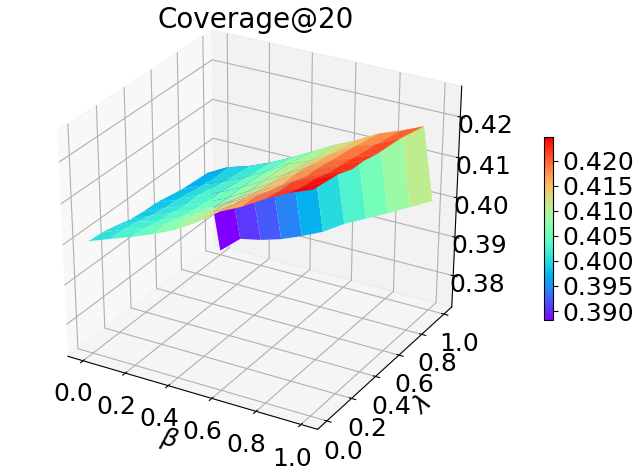}
\end{minipage}

\caption{\small{The impact of Hyper-Parameters $\lambda$ and $\beta$.}}
\label{fig:hyperparameter}
\end{figure*}

In this section, we conduct experiments to study the impact of hyper-parameter $\lambda$ and $\beta$ in DSM on RSC dataset, where both hyper-parameters are ranged in $[0,1]$. Figure~\ref{fig:hyperparameter-lambda}, Figure~\ref{fig:hyperparameter-beta} and Figure~\ref{fig:hyperparameter} present the impact of $\lambda$, $\beta$ and both of $\lambda$ and $\beta$ respectively. In these figures, from left to right are HR@20, MRR@20 and Coverage@20. The observations are summarized as follows:

\begin{itemize}
\item  As shown in Figure~\ref{fig:hyperparameter-lambda}, fixing $\beta=1$, the values of HR@20, MRR@20 and Coverage@20 are all decreased when we increase $\lambda$. Specifically, these metrics drop rapidly when $\lambda$ is larger than 0.5. The reason is that the impact of current session's length becomes larger when increasing $\lambda$, while that of compared session's length becomes less. However, the length of current session makes no sense, as discussed in Section~\ref{sec:algo:sim:adhc}, which leads to a trivial result.

\item If the value of $\beta$ is moderate, DSM leads to higher HR@20 and MRR@20. As shown in Figure~\ref{fig:hyperparameter-beta}, the values of HR@20 and MRR@20 are increasing as $\beta$ becomes larger from $0$, while declining when $\beta$ is greater than 0.5. In addition, the impact of item's popularity becomes greater as $\beta$ increases, as a result, the coverage@20 increases.

\item Figure~\ref{fig:hyperparameter} presents that DSM achieves the highest value in terms of HR@20 and MRR@20 when $\beta$ and $\lambda$ are all around 0.5. Under the same setting, DSM reaches a relative high value in terms of Coverage@20. Although DSM is able to obtain the highest value in terms of Coverage@20 when $\beta=1$, other metrics are poor. Therefore, we set $\mathbf {\beta=0.5}$ and $\mathbf {\lambda=0.5}$ in the following experiments to obtain both high accuracy and good diversity.
\end{itemize}

\subsection{Performance (RQ3)}\label{sec:expe:Performance}
The results of different approaches in terms of HR@20, MRR@20 and Coverage@20 on the three datasets is presented in Table~\ref{table:performance}. The following conclusions are observed:

\begin{itemize}
\item Due to ignoring the contextual information, IKNN achieves the worse performance in terms of HR@20 and MRR@20 on three datasets. Because the \textbf{limited accuracy of IKNN}, the best result in terms of Coverage@20 does not matter to the recommendation task.
\item Compared with the CKNN algorithms equipped with other similarity metrics and Original candidate selection method, \textbf{CKNN-DSM-Original achieves better accuracy} (i.e., HR@20 and MRR@20) when we set $\lambda=0.5$ and $\beta=0.5$. It is because DSM(0.5, 0.5) incorporates reasonable graph information when calculating the session similarities.
\item Compared with \textbf{Original} strategy, the proposed \textbf{EPCSR}, which focusing on both last click and other clicks, improves the performance on all three metrics. Specifically, CKNN-DSM(0.5, 0.5)-EPCSR outperforms CKNN-cosine-Original (which is the \textbf{state-of-the-art KNN} approach \cite{sessionknn}) by 7.4\%, 7.1\% and 10.6\% in terms of HR@20 (13.2\%, 8.8\% and 5.9\% in terms of MRR@20, 17.7\%, 5.5\% and 36.6\% in terms of Coverage@20) on RSC, RSCW and Atom datasets.
\end{itemize}
\begin{table*}[!t]
\centering
\small
\caption{\small{The performance on all datasets.}}\label{table:performance}
\begin{tabular}{ccccc}
\toprule
Datasets & Algorithms & HR@20  & MRR@20  & Coverage@20 \\
\midrule
\multirow{8}{*}{RSC}
&IKNN     & 0.5129&0.2051&\textbf{0.6267} \\ \cline{2-5}
&CKNN-cosine-Original    &0.6411 &0.2504 &0.3976 \\
&CKNN-HC-Original 		& 0.6422 &0.2513 &\textbf{0.4263} \\
&CKNN-MD-Original 		&0.6301 &0.2469 &0.3999 \\
&CKNN-MDHC-Original 		 &0.6393 &0.2497 &0.4229\\
&CKNN-DSM($0.5,0.5$)-Original & \textbf{0.6444}&\textbf{0.2515}&0.4099 \\\cline{2-5}
&CKNN-cosine-EPCSR & 0.6854&0.2815&0.4563 \\
&CKNN-DSM($0.5,0.5$)-EPCSR &\textbf{0.6888}&\textbf{0.2834}&\textbf{0.4678}\\\hline
\multirow{8}{*}{RSCW}
&IKNN     &0.4736 &	0.1975& 	\textbf{0.7590} \\ \cline{2-5}
&CKNN-cosine-Original  &0.6234 &0.2679 &0.5810  \\
&CKNN-HC-Original 		&0.6289 &0.2674 &\textbf{0.6151} \\
&CKNN-MD-Original 		&0.6238 &0.2645 &0.5884 \\
&CKNN-MDHC-Original 		&0.6296 &0.2665 &0.6077 \\
&CKNN-DSM($0.5,0.5$)-Original & \textbf{0.6329}&\textbf{0.2688}&0.5939\\\cline{2-5}
&CKNN-cosine-EPCSR & 0.6641&0.2900&0.6089\\
&CKNN-DSM($0.5,0.5$)-EPCSR &\textbf{0.6678}&\textbf{0.2915}&\textbf{0.6131} \\\hline
\multirow{8}{*}{Atom}
&IKNN     & 0.0260&0.0066&\textbf{0.6065} \\ \cline{2-5}
&CKNN-cosine-Original    &0.0568 &0.0068 &0.1509  \\
&CKNN-HC-Original 	 	 &0.0520 &0.0065 &\textbf{0.2357} \\
&CKNN-MD-Original 	 	&0.0546 &0.0069 &0.1767 \\
&CKNN-MDHC-Original 	  &0.0534 &0.0067 &0.2319 \\
&CKNN-DSM($0.5,0.5$)-Original & \textbf{0.0620}&\textbf{0.0070}&0.2006\\\cline{2-5}
&CKNN-cosine-EPCSR & 0.0572&0.0070&0.1546 \\
&CKNN-DSM($0.5,0.5$)-EPCSR &\textbf{0.0628}&\textbf{0.0072}&\textbf{0.2061}\\
\bottomrule
\end{tabular}
\end{table*}

\section{Conclusions}\label{sec:conclu}

In this paper, we propose a new contextual KNN approach for session-based recommendation, which incorporates the power of diffusion-based similarity method DSM and candidate selection method EPCSR. It gains performance improvement from these advantages: (1) adopting DSM, the session-item graph structure is utilized in the procedure of similarity calculation; (2) through guaranteeing the ratio of different clicks' recent relevant sessions in the recent session set of current session, EPCSR is able to capture the items that co-occurred with different historical clicked items in the same session efficiently.
We conducted extensive experiments on three benchmark datasets to compare the effectiveness of our approach and the state-of-the-art KNN approaches for session-based recommendation. Our experimental results demonstrate that our approach obtains better performance.

\bibliographystyle{ACM-Reference-Format}
\bibliography{paper}

\end{document}